\documentclass[a4wide,11pt]{article}
\usepackage{amsmath,amstext,amsfonts}
\def\bm#1{\mbox{\boldmath $#1$}}
\def\teigi{\stackrel{\rm def}{=}}

\makeatletter
  
  \@addtoreset{equation}{section}
\makeatother

\def\nyoro{\mathop{<}_{\sim}}
\begin{document}

\title{  Effective  QCD  Partition Function in Sectors with Non-Zero
  Topological Charge and 
  Itzykson-Zuber Type Integral }
\author{Toshinao A{\small KUZAWA}\thanks{akuzawa@monet.phys.s.u-tokyo.ac.jp} 
{\small and}  Miki W{\small   ADATI}\\
{\small\it  
Department of Physics, Graduate School of Science, University of
  Tokyo,}\\ {\small\it Hongo 7-3-1, Bunkyo-ku, Tokyo, 113}}
\maketitle
\abstract{It was 
conjectured by Jackson {\it et.al.} that the finite volume effective
partition 
  function of   QCD with the topological charge  $M-N$ coincides with  the
  Itzyskon-Zuber type integral for $M\times N$ rectangular
  matrices. In the present article we give 
   a proof of this conjecture, in which  the original 
  Itzykson-Zuber   integral is utilized. 
}

\section{Introduction}
Let us  start with  
quantum chromodynamics 
 (QCD) with $N$ flavors and 
a nonzero vacuum angle $\theta$. 
Assume that the quarks have masses $m_1,m_2,\cdots,m_N$. 
 We introduce the quark-mass matrix   ${\bm M}$ by
\begin{eqnarray}
  \label{vin1.01}
{\bm M}={\rm diag}(m_1,\cdots,m_N)~.  
\end{eqnarray}
Using matrices    $\gamma_\mu$, $\mu=1,2,3,4$,  defined by 
\begin{eqnarray}
  \label{vin0.1}
  \gamma_4={\rm i}~{\rm diag}(1,1,-1,-1)
\end{eqnarray}
and 
\begin{eqnarray}
  \label{vin0.2}
  \{\gamma_\mu,\gamma_\nu\}=-2\delta_{\mu\nu}~, 
\end{eqnarray}
 the Lagrangian  density           
in the Euclidean space is expressed as    
\cite{leutwyler-smilga1} 
\begin{eqnarray}
  \label{vin1}
  L=\frac{1}{4g^2}
\sum_{\mu,\nu=1}^N\sum_{a=1}^{N^2-1}
F_{\mu \nu}^aF_{\mu \nu}^a-{\rm i}\theta \omega
-{\rm i}{\bar q}\sum_{\mu=1}^{4}\gamma_{\mu}(\partial_{\mu}+{\rm
  i}A_{\mu})q 
+{\bar  q}_R{\bm M}q_L+{\bar  q}_L{\bm M}^{\dagger}q_R~,
\end{eqnarray}
where 
\begin{eqnarray}
  \label{vin2}
  \omega=\frac{1}{32\pi^2}{\rm
    tr}\epsilon^{\mu\nu\alpha\beta}F_{\mu\nu}{F}_{\alpha\beta}~ . 
\end{eqnarray}
We assume that the space-time constitutes the  four-dimensional torus
with the  volume $V$.  
The winding number (or topological charge) $\nu$ of the gauge field,
which is  expressed as  
\begin{eqnarray}
  \label{vin3}
\nu =   \int_V d^4x \omega(x)~,
\end{eqnarray}
is a topological invariant. 
 The partition function ${\cal Z}(\theta)$ is written as 
\begin{eqnarray}
  \label{vin4}
  {\cal Z}(\theta)=\sum_{\nu=-\infty}^{\infty}e^{{\rm i}\nu \theta}
{\cal Z}_{\nu}~,
\end{eqnarray}
where ${\cal Z}_{\nu}$ is the partition function in the sector with
topological  
charge $\nu$. 
We consider  the region in which 
the volume $V$ is large and 
\begin{eqnarray}
  \label{vin5}
V\sigma {\cal S}\nyoro 1~,  
\end{eqnarray}
where $\sigma$ and ${\cal S}$ represent respectively  the quark
condensate and the 
quark mass scale.   
In the low-energy theory, the Goldstone bosons dominate
the partition   function. 
In the present case 
the
Goldstone bosons are described by matrix fields $U(x)$ in $SU(N)$.   The
partition function takes the form
\begin{eqnarray}
  \label{vin6}
  {\cal Z}=\int_{U(x)\in SU(N)}{\delta U}\exp\left\{
-\int_{V}d^4x\left[
\frac{f^2}{4}{\rm tr}(\partial_{\mu}U^{\dagger}\partial_{\mu}U)-
\sigma {\rm Re}\left(
\exp({\rm i}\theta/N){\rm tr}{\bm M}U^{\dagger}
\right)+\cdots
\right]
\right\}~,
\end{eqnarray}
where $f$ is the pion-decay constant
and $\rm Re$ means the real part. 
  By setting $U(x)=U_0 U_1(x)$
and ignoring the non-zero modes fluctuations associated with $U_1(x)$, 
we arrive at the effective QCD partition function in terms of the collective
variable $U_0$ \cite{leutwyler-smilga1},   
\begin{eqnarray}
  \label{vin7}
 {\cal  Z}=\int_{SU(N)} d\mu(U_0)
\exp
\left\{
V\sigma {\rm Re}\left(
{\rm e}^{{\rm i}\theta/N}{\rm tr}({\bm M}U_0^{\dagger})
\right)
\right\}~.
\end{eqnarray}
Hereafter we denote by $\mu$ the normalized measure with the two-sided
invariance. 
From (\ref{vin7}) it is deduced that  ${\cal Z}_{\nu}$  is  expressed
as an integral over 
$U(N)$: 
 \begin{eqnarray}
  \label{vin8}
{\cal Z}_{\nu}=  \int_{U(N)} d\mu (U)(\det U)^{\nu}\exp(V\sigma{\rm
  Re~ tr }{\bm M}U)~. 
\end{eqnarray}
In case  quarks have equal  masses, 
the $U(N)$-integral in (\ref{vin8}) is carried out in
\cite{leutwyler-smilga1}.  For different quark masses the explicit
expression for ${\cal Z}_{\nu=0}$ is obtained in
\cite{jackson-sener-verbaarschot1}. 
The key of the derivation in \cite{jackson-sener-verbaarschot1} is
the Itzykson-Zuber type 
 integral for  complex matrices
\begin{eqnarray}
  \label{va1}
\lefteqn{
  \int_{U(M)} d\mu(U)\int_{U(N)} d\mu (U')\exp
\left(
{\rm Re}~ {\rm tr}(U'\Psi U \Xi)
\right)}\nonumber\\
&&=\frac{2^{N(M-1)}}{N!}
\prod_{j=0}^{N}(1+j)!(M-N+j)!
\prod_{i=1}^N(x_iy_i)^{\frac{1}{2}}
\frac{\det\{I_{M-N}({x_k y_l })\}_{k,l}}{\xi_0(x)\xi_0(y)}
~,
\end{eqnarray}
where 
\begin{eqnarray}
  \label{va3}
  \Xi=
\left(
\begin{array}{cccc}
x_1 & 0   & \ldots & 0                  \\
0   & x_2 & \ldots & 0              \\
\multicolumn{4}{c}{\dotfill}\\
\multicolumn{4}{c}{\dotfill}\\
0   & 0   & \ldots & x_N \\
\multicolumn{4}{c}{\dotfill}\\
0 & \multicolumn{2}{c}{\dotfill} & 0 
  \end{array}
\right)~,   
\end{eqnarray}
\begin{eqnarray}
  \label{vin10}
 \Psi &=&
\left(
\begin{array}{cccccc}
y_1 & 0   & \ldots & 0      & \ldots & 0                  \\
0   & y_2 & \ldots & 0      & \ldots & 0                   \\
\multicolumn{6}{c}{\dotfill}\\
\multicolumn{6}{c}{\dotfill}\\
0   & 0   & \ldots & y_{N} & \ldots & 0 \\
  \end{array}
\right)~,
\end{eqnarray}
and $\xi_0(x)$ is defined by 
\begin{eqnarray}
  \label{9.1a}
  \xi_0(x)=
\prod_{i,j=1}^N(x_i^2-x_j^2)\prod_{i=1}^N
  x_i^{M-N+\frac{1}{2}}~.  
\end{eqnarray}
There is a straightforward
 correspondence between ${\cal Z}_{\nu=0}$ and (\ref{va1}) for $M=N$. What
 was conjectured in \cite{jackson-sener-verbaarschot1} is that
 ${\cal Z}_{\nu}$ for general $\nu$ also corresponds to (\ref{va1}) for
 $M=N+\nu$. In the present  article we shall give a proof to this
 conjecture. In our proof we make use of the original Itzykson-Zuber
 integral\cite{itzykson-zuber1}. 
\section{Correspondence between Itzykson-Zuber type Integral and
  Partition Function} 
The correspondence between ${\cal Z}_{\nu}$ and the integral
(\ref{va1}) is easily 
certified. We assume that $M\ge N$ without losing any generality. 
We
 set  
\begin{eqnarray}
  \label{va2}
  \Psi =v {\cal M}~,
\end{eqnarray}
where $v=V\sigma$  is a real positive number and ${\cal M}$ is an 
$N\times M$ matrix such as 
\begin{eqnarray}
  \label{13.2a}
{\cal M}&=&
\left(
\begin{array}{cccccc}
m_1 & 0   & \ldots & 0      & \ldots & 0                  \\
0   & m_2 & \ldots & 0      & \ldots & 0                   \\
\multicolumn{6}{c}{\dotfill}\\
\multicolumn{6}{c}{\dotfill}\\
0   & 0   & \ldots & m_{N} & \ldots & 0 \\
  \end{array}
\right)~.  
\end{eqnarray}
Set also
\begin{eqnarray}
  \label{va3}
  \Xi={\bm 1}_{M,N}\equiv 
\left(
\begin{array}{cccc}
1 & 0   & \ldots & 0                  \\
0   & 1 & \ldots & 0              \\
\multicolumn{4}{c}{\dotfill}\\
\multicolumn{4}{c}{\dotfill}\\
0   & 0   & \ldots & 1 \\
\multicolumn{4}{c}{\dotfill}\\
0 & \multicolumn{2}{c}{\dotfill} & 0 
  \end{array}
\right)~.  
\end{eqnarray}
Then the Itzykson-Zuber type integral 
(\ref{va1}) becomes 
\begin{eqnarray}
  \label{va4}
  G_{M,N}({\cal M})\teigi
 \int_{U(M)} d\mu(U)\int_{U(N)} d\mu (U')\exp
\left(
{\rm Re}~ v{\rm tr}(U' {\cal M} U {\bm 1}_{M,N})
\right)~.
\end{eqnarray}
In case $M=N$ the double integral in (\ref{va4}) reduces to a single
integral: 
\begin{eqnarray}
  \label{va4.01}
    G_{N,N}({\cal M})=
 \int_{U(N)} d\mu^M(U)\exp
\left(
{\rm Re}~ v{\rm tr}( {\cal M} U )
\right)~.
\end{eqnarray}
This is exactly the partition function ${\cal Z}_{0}$ in the sector with
topological 
charge $\nu=0$.
Furthermore,   by giving examples 
 which suggest 
the existence of the  correspondence for general $\nu$, 
it was   conjectured in  \cite{jackson-sener-verbaarschot1}  that 
\begin{eqnarray}
  \label{vb5}
   G_{M,N}({\cal M})={\sf constant}\times {\cal Z}_{M-N}~.
\end{eqnarray}
\section{Proof of the Conjecture}

%
For definiteness sake 
we shall add a suffix which specifies the number of the flavors
to   the partition function in the sector with topological
charge $\nu$, that is,      
\begin{eqnarray}
  \label{vb1}
{\cal Z}^N_{\nu}=  \int_{U(N)} d\mu (U)(\det U)^{\nu}\exp(v{\rm Re~ tr }MU)~.
\end{eqnarray}
 By properly choosing   $S\in SU(N)$
and $\Lambda={\rm diag}(e^{{\rm i}x_1},e^{{\rm i}x_2},\cdots,e^{{\rm
    i}x_N})$, where $x_i\in {\Bbb R}$, 
 the  matrix $U\in U(N)$ is expressed as 
\begin{eqnarray}
  \label{vb1.01}
  U=S\Lambda S^{\dagger}~.
\end{eqnarray}
The well-known transformation of the variables of the integrations
\cite{mehta1} reads 
\begin{eqnarray}
  \label{vb1.02}
  \int_{U(N)}d\mu(U)=\frac{1}{N!(2\pi)^N}
\int_{SU(N)}d\mu(S)\left(\prod_{i=1}^N\int_0^{2\pi}dx_i\right)
\prod_{i<j}\sin^2\frac{x_i-x_j}{2}    ~.
\end{eqnarray}
In terms of $S$ and $\Lambda$,  
(\ref{vb1}) is  transformed to
\begin{eqnarray}
  \label{vb2}
  {\cal Z}^N_{\nu}= \frac{1}{N!(2\pi)^N}
\prod_{i=1}^N \left(\int_{0}^{2\pi}dx_i  {\rm e}^{{\rm
        i}x_i}\right) 
\prod_{i<j}\sin^2\frac{x_i-x_j}{2}
\int_{SU(N)}
d\mu (S)\exp(v{\rm  tr }MS\Lambda S^{\dagger})~.
\end{eqnarray}
The expression (\ref{vb2}) has a great advantage because 
 the $SU(N)$ integration is readily carried out by using the original 
Itzykson-Zuber integral\cite{itzykson-zuber1}, 
\begin{eqnarray}
  \label{vb3}
  I(A,B;\beta)&=&\int_{SU(N)} d\mu(U)\exp[\beta{\rm tr}AUBU^{\dagger}]\\
&=&\beta^{-N(N-1)/2}\prod_{n=1}^{N-1}n!
\frac{\det \left\{{\rm e}^{\beta a_i b_j}\right\}_{1\le i,j\le N}
}{\Delta(a)\Delta(b)}~,
\end{eqnarray}
where $\Delta$ is the Vandermonde determinant, for instance, 
\begin{eqnarray}
  \label{van1}
  \Delta(a)=\prod_{1\le i< j\le N}(a_i-a_j)~.
\end{eqnarray}
Consequently 
 there remain  the integrations over $N$ variables: 
\begin{eqnarray}
  \label{vb4}
 {\cal  Z}^N_{\nu}= v^{-N(N-1)/2}\prod_{n=1}^{N-1}n!
\prod_{i=1}^N \left(\int_{0}^{2\pi}dx_i  {\rm e}^{{\rm i}\nu x_i}\right)
\prod_{i<j}\sin^2\frac{x_i-x_j}{2}
\frac{\det\left\{ {\rm e}^{v m_j\cos x_i }\right\}_{1\le i,j\le
    N}}{\Delta(\cos 
  x)\Delta(m)}~. 
\end{eqnarray}
In the next place, let us examine $  G_{M,N}({\cal M})$. 
It can be shown that 
\begin{eqnarray}
  \label{va5}
  G_{M,N}({\cal M})&=&  G_{M,M}({\tilde {\cal M}})
=\int d\mu^M(U)\exp
\left(
{\rm Re}~ v{\rm tr}( {\tilde {\cal M}} U )
\right)
~,
\end{eqnarray}
where 
\begin{eqnarray}
  \label{va6}
  {\tilde{\cal M}}&=&{\rm diag}(m_1,m_2,\cdots,m_N,0,\cdots,0)~.
\end{eqnarray}
For the right-hand-side of (\ref{va5}) a similar kind of
decomposition as for ${\cal Z}^N_{\nu}$ is possible. 
As a result,  $G_{M,N}({\cal M})$ is rewritten as 
\begin{eqnarray}
  \label{vb4.5}
  G_{M,N}({\cal M})=\frac{v^{-M(M-1)/2}}{M!(2\pi)^M}
\prod_{n=1}^{M-1}n!
\prod_{i=1}^M \left(\int_{0}^{2\pi}dx_i  \right)
\prod_{i<j}^{M}\sin^2\frac{x_i-x_j}{2}
\frac{\det \left\{{\rm e}^{v {\tilde m}_j\cos x_i }\right\}_{1\le i,j
    \le M}
}{\Delta(\cos x)\Delta({\tilde m})}~,
\end{eqnarray}
where 
\begin{eqnarray}
  \label{vb4.6}
  {\tilde m_i}=\left\{
  \begin{array}{cc}
m_i &(1\le i\le N)\\
0 & (N+1\le i \le M)
  \end{array}
\right. ~.
\end{eqnarray}
We shall prove the conjecture that 
\begin{eqnarray}
  \label{vb5}
   G_{M,N}({\cal M})={\sf constant}\times {\cal Z}^N_{M-N}~,
\end{eqnarray}
by comparing (\ref{vb4}) with (\ref{vb4.5}). 

We start from (\ref{vb4.5}). 
The  fraction  
\begin{eqnarray}
  \label{vb5.9}
   \frac{\det \left\{{\rm e}^{v {\tilde m}_j\cos x_i }\right\}_{1\le
       i,j \le M}
}{\Delta(\cos x)\Delta({\tilde m})}
\end{eqnarray}
in (\ref{vb4.5}) is decomposed by using the Schur function
$S_{\{\lambda\}}$ as  
\begin{eqnarray}
  \label{vb6}
&=&
\prod_{i=1}^M\left(
\sum_{k_i=0}^{\infty}
\frac{v^{k_i}}{k_i!}
\right)
\frac{\prod_{j=1}^M \cos^{k_j} x_j}{\Delta(\cos x)}
 \frac{\det { {\tilde m}_j^{k_i} }}{\Delta({\tilde m})}\nonumber\\
&=&
\sum_{k_M=0}^{\infty}
\sum_{k_{M-1}(> k_M)}^{\infty}
\cdots 
\sum_{k_1(> k_2)}^{\infty}
\prod_{i=1}^M\left(
\frac{v^{k_i}}{k_i!}
\right)
\frac{\det\cos^{k_i} x_j}{\Delta(\cos x)}
 \frac{\det { {\tilde m}_j^{k_i} }}{\Delta({\tilde m})}\nonumber\\
&=&
\sum_{\lambda_M=0}^{\infty}
\sum_{\lambda_{M-1}(\ge \lambda_M)}^{\infty}
\cdots 
\sum_{\lambda_1(\ge \lambda_2)}^{\infty}
\prod_{i=1}^M\left(
\frac{v^{\lambda_i+M-i}}{(\lambda_i+M-i)!}
\right)
S_{\{\lambda\}}(\cos x)
S_{\{\lambda\}}({\tilde m})~.
\end{eqnarray}
For our purpose,  it is important to note that 
  the Schur functions have the property\cite{macdonald1}, 
\begin{eqnarray}
  S_{\{\lambda\}}({\tilde
    m})&=&S_{\{\lambda\}}(m_1,\cdots,m_N,0,\cdots,0)\nonumber\\
  \label{vb7}
&=&\left\{
    \begin{array}{lcl}
  S_{\{\lambda_1,\cdots,\lambda_N\}}(m_1,\cdots,m_N)
&:& \mbox{\rm   if $\lambda_{N+1}=\lambda_{N+2}=\cdots =\lambda_M =0$}\\
0 &:& \mbox{\rm otherwise}
    \end{array}
\right. ~.
\end{eqnarray}
It follows from (\ref{vb7}) that  
\begin{eqnarray}
  \label{vb8}
 \frac{\det\left\{ {\rm e}^{v {\tilde m}_j\cos x_i }\right\}_{1\le i,j 
     \le M}
}{\Delta(\cos x)\Delta({\tilde m})}
&=&  
\prod_{i=1}^{M-N}
\frac{v^{i-1}}{(i-1)!}
\sum_{\lambda_N=0}^{\infty}
\sum_{\lambda_{N-1}(\ge \lambda_N)}^{\infty}
\cdots 
\sum_{\lambda_1(\ge \lambda_2)}^{\infty}\nonumber\\
&&\times
\prod_{i=1}^N\left(
\frac{v^{\lambda_i+M-i}}{(\lambda_i+M-i)!}
\right)
S_{\{\lambda_1,\cdots,\lambda_N,0\cdots,0\}}(\cos x)
S_{\{\lambda_1,\cdots,\lambda_N\}}(m)
~.
\end{eqnarray}
Since the notational convention of the Schur function is rather inconvenient 
to the discussions below, 
 we introduce a function $T^{(L)}_{\{k_1,\cdots,k_L\}}$  defined by
\begin{eqnarray}
  \label{vb8.1}
  T^{(L)}_{\{k_1,\cdots,k_L\}}(x_1,\cdots,x_L)=\frac{\det\{x_i^{k_j}\}_{1\le i,j\le L}} 
 {\Delta(x) } ~.
\end{eqnarray}
The Schur function and $T^{(L)}_{\{k_1,\cdots,k_L\}}$ are  related by 
\begin{eqnarray}
  \label{vb8.2}
S_{\{\lambda_1,\cdots,\lambda_L\}}=
T^{(L)}_{\{k_1+L-1,k_2+L-2,\cdots,k_{L-1}+1, k_L\}}~.
\end{eqnarray}
In terms of $T^{(L)}_{\{k_1,\cdots,k_L\}}(x_1,\cdots,x_L)$,  
(\ref{vb8}) is  rewritten as 
\begin{eqnarray}
  \label{vb8.3}
   \frac{\det \left\{{\rm e}^{v {\tilde m}_j\cos x_i }\right\}_{1 \le
       i,j\le M}
}{\Delta(\cos x)\Delta({\tilde m})}
&=&  
\prod_{i=1}^{M-N}
\frac{v^{i-1}}{(i-1)!}
\sum_{\lambda_N=0}^{\infty}
\sum_{\lambda_{N-1}(\ge \lambda_N)}^{\infty}
\cdots 
\sum_{\lambda_1(\ge \lambda_2)}^{\infty}\nonumber\\
&\times&
\prod_{i=1}^N\left(
\frac{v^{\lambda_i+M-i}}{(\lambda_i+M-i)!}
\right)
T^{(M)}_{\{\lambda_1+M-1,\cdots,\lambda_N+M-N,M-N-1,\cdots,1,0\}}(\cos x)
T^{(N)}_{\{\lambda_1+N-1,\cdots,\lambda_N\}}(m)\nonumber\\
&=&  
\prod_{i=1}^{M-N}
\frac{v^{i-1}}{(i-1)!}
\sum_{k_N=0}^{\infty}
\sum_{k_{N-1}(\ge k_N)}^{\infty}
\cdots 
\sum_{k_1(\ge k_2)}^{\infty}\nonumber\\
&&\times
\prod_{i=1}^N\left(
\frac{v^{k_i}}{k_i!}
m_i^{N-M}
\right)
T^{(M)}_{\{k_1,\cdots,k_N,M-N-1,\cdots,1,0\}}(\cos x)
T^{(N)}_{\{k_1,\cdots,k_N\}}(m)
~.
\end{eqnarray}
Note that  
\begin{eqnarray}
  \label{vb9}
  T^{(M)}_{\{k_1,\cdots,k_N,M-N-1,M-N-2,\cdots,1,0\}}(\cos x)
=\frac{1}{\Delta(\cos x)}\sum_{P\in \sigma_M}(-1)^P
\prod_{i=1}^N \cos^{k_{i}}x_{P_i} 
\prod_{\alpha=1}^{M-N}\cos^{M-N-i}x_{P_{N+\alpha}} ~,
\end{eqnarray}
where $\sigma_M$ is the permutation of $\{1,2,\cdots,M\}$. 
By introducing  $\sigma^N_M$ by
\begin{eqnarray}
  \label{vb10}
  \sigma^N_M=\left\{
P\in \sigma_M:
P_{N+1}>P_{N+2}\cdots >P_{M},~P_{1}>P_{2}\cdots >P_{N}
\right\}~,
\end{eqnarray}
 we may express the left hand side of (\ref{vb9}) 
 in terms of  $T^{(N)}_{\{k_1,\cdots,k_N\}}$ as 
\begin{eqnarray}
  \label{vb11}
&&  T^{(M)}_{\{k_1,\cdots,k_N,M-N-1,M-N-2,\cdots,1,0\}}(\cos x) 
\nonumber\\
&&=\frac{1}{\Delta(\cos x)}\sum_{P\in \sigma^N_M}(-1)^P
\det\left\{  \cos^{k_{i}}x_{P_j}  \right\}_{1\le i,j\le N}
\det\left\{\cos^{M-N-i}x_{P_{N+j}} \right\}_{1\le i,j\le
  M-N}\nonumber\\
&&=\sum_{P\in \sigma^N_M}
T^{(N)}_{\{k_1,\cdots,k_N\}}
(\cos x_{P_1},\cos x_{P_2},\cdots,\cos x_{P_N})
\prod_{i=1}^N\prod_{j=N+1}^M 
\frac{1}{\cos x_{P_i}-\cos x_{P_j}}~.
\end{eqnarray}
Putting (\ref{vb4.5}), (\ref{vb8.3}) and (\ref{vb11})
together we get 
\begin{eqnarray}
  \label{vb12}
&&  G_{M,N}({\cal M})=\frac{
v^{N(N-2M+1)/2}
}{M!(2\pi)^M}
\prod_{n=M-N}^{M-1}n!
\sum_{k_N=0}^{\infty}
\sum_{k_{N-1}(\ge k_N)}^{\infty}
\cdots 
\sum_{k_1(\ge k_2)}^{\infty}
\prod_{i=1}^N\left(
\frac{v^{k_i}}{k_i!}m_i^{N-M}
\right)
T^{(N)}_{\{k_1,\cdots,k_N\}}(m)
\nonumber\\
&&\times
\prod_{i=1}^M \left(\int_{0}^{2\pi}dx_i  \right)
\prod_{i<j}^{M}\sin^2\frac{x_i-x_j}{2}
\sum_{P\in \sigma^N_M}
T^{(N)}_{\{k_1,\cdots,k_N
  \}}(\cos x_{P_1},\cos x_{P_2},\cdots,\cos x_{P_N})
\nonumber\\
&&\times \prod_{i=1}^N\prod_{\alpha=N+1}^M 
\frac{1}{\cos x_{P_i}-\cos x_{P_{\alpha}}}\nonumber\\
&&
=\frac{
v^{N(N-2M+1)/2}
}{M!(2\pi)^M}
{}_M{\rm C}_N\prod_{n=M-N}^{M-1}n!
\sum_{k_1\ge k_2  \ge \cdots k_N}
\prod_{i=1}^N\left(
\frac{v^{k_i}}{k_i!}m_i^{N-M}
\right)
T^{(N)}_{\{k_1,\cdots,k_N\}}(m)
\nonumber\\
&&\times
\prod_{i=1}^N \left(\int_{0}^{2\pi}dx_i  \right)
\prod_{i<j}^{N}\sin^2\frac{x_i-x_j}{2}
T^{(N)}_{\{k_1,\cdots,k_N
  \}}(\cos x_{1},\cos x_{2},\cdots,\cos x_{N})\nonumber\\
&&\times 
\prod_{i=N+1}^M \left(\int_{0+i\epsilon}^{2\pi+i\epsilon}dx_i  \right)
\prod_{N+1\le i<j\le M}\sin^2\frac{x_i-x_j}{2}
 \prod_{i=1}^N\prod_{\alpha=N+1}^M
\frac{\sin^2\frac{x_i-x_{\alpha}}{2}}{\cos x_{i}-\cos x_{\alpha}}
~.
\end{eqnarray}
The infinitesimal positive
number $\epsilon$ is introduced to avoid the ``pseudo'' poles at
$x_i=-x_{\alpha}$. 
With a change of variables $y_i={\rm e}^{{\rm i} x_i}$, 
the integration over $\{x_{i}:N+1\le i \le M\}$ in the last expression 
in (\ref{vb12}) is transformed to the contour integral as 
\begin{eqnarray}
  \label{vb13}
&&2^{-(M-N)(M-1)} (-1)^{(M-N)(M-N-1)/2}
\prod_{i=N+1}^M \left(-{\rm
    i}\oint_{C_{\epsilon}}\frac{dy_i}{y_i}  \right) \nonumber\\
&&\times
\prod_{N+1\le i<j\le M}(y_i-y_j)(y_i^{-1}-y_j^{-1})
 \prod_{i=1}^N\prod_{\alpha=N+1}^M
\frac{y_{\alpha}-y_{i}}{y_{\alpha}y_{i}-1}
\end{eqnarray}
where  $C_{\epsilon}$ is a circle  with radius ${\rm e}^{-\epsilon}$
centered at the origin of the 
complex plane.  
 Equation 
(\ref{vb13}) is immediately integrated out:
\begin{eqnarray}
  \label{vb14}
  =2^{-(M-N)(M-1)}(2\pi)^{M-N} (-1)^{(M-N)(M+N-1)/2}(M-N)!
 \prod_{i=1}^Ny_{i}^{M-N}~.
\end{eqnarray}
Inserting (\ref{vb14}) in (\ref{vb12}) 
we find that 
\begin{eqnarray}
  \label{vb15}
  &&  G_{M,N}({\cal M})=
  C_{(M,N)}
\sum_{k_1\ge k_2  \ge \cdots k_N}
\prod_{i=1}^N\left(
\frac{v^{k_i}}{k_i!}
m_i^{N-M}
\right)
T^{(N)}_{\{k_1,\cdots,k_N\}}(m)
\nonumber\\
&&\times
\prod_{i=1}^N \left(\int_{0}^{2\pi}dx_i  {\rm e}^{{\rm i}(M-N)x_i} \right)
\prod_{i<j}^{N}\sin^2\frac{x_i-x_j}{2}
T^{(N)}_{\{k_1,\cdots,k_N
  \}}(\cos x_{1},\cos x_{2},\cdots,\cos x_{N})~,
\end{eqnarray}
where
\begin{eqnarray}
  \label{vb16}
  C_{(M,N)}=v^{N(N-2M+1)/2}
2^{-(M-N)(M-1)}(2\pi)^{M-N}
  (-1)^{(M-N)(M+N-1)/2}\frac{1 }{N!}
\prod_{n=M-N}^{M-1}n!~.
\end{eqnarray}
The summation in  (\ref{vb15}) reads 
\begin{eqnarray}
  \label{vb17}
 && \sum_{k_1\ge k_2  \ge \cdots k_N}
\prod_{i=1}^N\left(
\frac{v^{k_i}}{k_i!}
m_i^{N-M}
\right)
T^{(N)}_{\{k_1,\cdots,k_N\}}(m)
T^{(N)}_{\{k_1,\cdots,k_N
  \}}(\cos x_{1},\cos x_{2},\cdots,\cos x_{N})\nonumber\\
&&=\frac{1}{\Delta(\cos x)\Delta(m)}
\sum_{k_1\ge k_2  \ge \cdots k_N}
\prod_{i=1}^N\left(
\frac{v^{k_i}}{k_i!}m_i^{N-M}
\right)
\det\big\{m_i^{k_{j}}\big\}_{1\le i,j\le N}
\det\big\{\cos^{k_{j}} x_i
\big\}_{1\le i,j\le N}\nonumber\\
&&=\frac{1}{\Delta(\cos x)\Delta(m)}\prod_{i=1}^N m_i^{N-M}
\det\bigg\{\exp(vm_i \cos x_j)
\bigg\}_{1\le i,j\le N}~.
\end{eqnarray}
Finally by combining (\ref{vb15}) and (\ref{vb17}) we obtain 
\begin{eqnarray}
  \label{vb20}
  G_{M,N}({\cal M})&=&
  C_{(M,N)}\prod_{i=1}^N m_i^{N-M}
\prod_{i=1}^N \left(\int_{0}^{2\pi}dx_i  {\rm e}^{{\rm i}(M-N)x_i} \right)
\nonumber\\
&&\times 
\prod_{i<j}^{N}\sin^2\frac{x_i-x_j}{2}
\frac{1}{\Delta(\cos x)\Delta(m)}
\det\bigg\{\exp(vm_i \cos x_j)
\bigg\}_{1\le i,j\le N}~,
\end{eqnarray}
which is proportional to (\ref{vb4}) with $\nu=M-N$. This 
proves the conjecture (\ref{vb5}). 
\section{Discussions}
The partition function (\ref{vin8}) with $\nu=0$ has the two-sided $U(N)$
invariance.   
That makes it possible to carrry out the integration. 
If $\nu\ne 0$,  the partition function is invariant under $SU(N)$ but
not invariant under $U(N)$. That is, $U(1)$  symmetry is
broken. Despite an 
impression that the remaining  $SU(N)$ symmetry is still large, it is
not easy to treat the partition function analytically. 

It is known \cite{akuzawa5}     that the Itzykson-Zuber type integral
for rectangular 
matrices is derived  by considering the zero-curvature limit of the
isotropic diffusion on 
the cosets $SU(M+N)/S(U(M)\times U(N))$ and
$SU(M,N)/S(U(M)\times U(N))$. These cosets are classified as 
the Riemannian symmetric space of type A${\rm
  I\!I\!I}$ in the mathematical context\cite{helgason1}.  
We have proved in the present article 
 the conjecture (\ref{vb5}) that the partition functions in the sectors 
 with non-zero topological charges equals the Itzykson-Zuber type
 integral for rectangular matrices up to the normalization.  
This 
correspondence makes it possible to  apply a full knowledge of
the symmetric 
spaces 
to the analysis of the
sectors with 
$\nu\ne0$. 

\begin{thebibliography}{99}

\bibitem{leutwyler-smilga1}
H.Leutwyler and A.Smilga, Spectrum of Dirac operator and role of winding number
  in QCD. {\em Phys.Rev.D\/}, {\bf 46}(12), 5607 (1992).

\bibitem{jackson-sener-verbaarschot1}
A.~Jackson, M.~\c{S}ener, and J.~Verbaarschot, Finite volume partition
  functions and Itzykson-Zuber integrals. {\em Phys.Lett.B\/}, {\bf 387}, 355
  (1996).

\bibitem{itzykson-zuber1}
C.Itzykson and J.-B.Zuber, The planar approximation. II. {\em J.Math.Phys.},
  {\bf 21}, 411 (1980).

\bibitem{mehta1}
M.L.Mehta, {\em Random Matrices\/}. Academic Press, second edn. (1991).

\bibitem{macdonald1}
I.G.Macdonald, {\em Symmetric Functions and Hall Polynomials\/}. Oxford
  University Press, Oxford, second edn. (1995).

\bibitem{akuzawa5}
T.Akuzawa and M.Wadati, Diffusions on symmetric spaces of type A${\rm I\!I\!I}$
  and random matrix theories for rectangular matrices. {\em J.Phys.A\/}, {\bf
  31}, 1713 (1998).

\bibitem{helgason1}
S.Helgason, {\em Differential Geometry, Lie Groups and Symmetric Spaces\/}.
  Academic Press, New York (1978).

\end{thebibliography}

\end{document}